# Hypercomputation, Frege, Deleuze: Solving Thomson's Lamp

**ABSTRACT**

We present the first known solution to the original supertask, the Thomson Lamp Paradox. Supertasks, or infinite operations, form an integral part of the emerging study of hypercomputation (computation in excess of classical Turing machine capabilities) and are of philosophical interest for the paradoxes they engender.

We also offer preliminary resources for classifying computational complexity of various supertasks. In so doing we consider a newly apparent paradox between the metrical limit and the ordinal limit. We use this distinction between the metrical and ordinal limits to explain the shortcomings both of Thomson's original formulation of the Lamp Paradox and Benacerraf's consequent critique.

We resolve this paradox through a careful consideration of transfinite ordinals and locate its ambiguity as inherent to the identity relation under logic with a close reading of Frege's *Begriffsschrift*. With this close reading in hand we expose how the identity relation is counter-intuitively polyvalent and, with supertasks, how the logico-mathematical field operates on the basis of Deleuzian point-folds. Our results combine resources from philosophy, mathematics, and computer science to ground the field of hypercomputation for logically rigorous study.

**INTRODUCTION**

In "Tasks and Super-Tasks" Thomson proposes a number of supertasks (infinite operations) and each time asks the same question: *Can we determine what the state of the world is after the supertask's completion?* The paradoxes therein convinced Thomson and other philosophers that any supertask was self-contradictory.

It was not until Benacerraf demonstrated a logical error in Thomson's paradox proof roughly ten years later that the topic of supertasks opened once more. However, Benacerraf stops short of answering Thomson's thematic question: Do we have a way of explaining what it means for a supertask to be finished?

We present an original solution to Thomson's most celebrated paradox, the Thomson Lamp, by answering this question in the affirmative. We begin by discussing two other supertask paradoxes Thomson supplies as way of mathematical and philosophical preparation. We discuss resources for categorizing the complexity of supertasks and consider an additional paradox between metrical and ordinal limits uncovered in our solution to Thomson's Lamp. We conclude with a close reading of the identity relation in Frege's *Begriffsschrift* and a provocative suggestion of the supertask as a Deleuzian point-fold.

**SUPERTASKS AND PARADOX**

We owe Thomson a great deal for succinctly formulating the main question of supertasks. "Is it conceivable that someone should have completed an infinite number of tasks? Do we know what this would be like?" (2).[1] Thomson takes as his inspiration the paradoxes given by Zeno of Elea and offers several of his own to circumscribe the question.

We consider three main paradoxes Thomson offers: infinitely divisible chocolate, a man running from [0, 1), and the Thomson Lamp. From the outset Thomson is precise about what qualifies as a supertask. Immediately following the quotation above, Thomson writes "It is necessary here to avoid a common confusion…to say that some operation can be performed infinitely often is not to say that a super-operation can be performed" (2).

Thomson distinguishes between the capacity to perform a single operation an infinite number of times and the capacity to perform a single, infinite operation. "Suppose (A) that every lump of chocolate can be cut in two, and (B) that the result of cutting a lump of chocolate in two is always that you get two lumps of chocolate. It follows that every lump of chocolate is infinitely divisible" (2).

From this chocolate splitting he derives a means to differentiate between a set of infinitely many possibilities and the possibility of infinity itself (3). "But to say that a lump is infinitely divisible is just to say that it can be cut into any number of parts. Since there is an infinite number of numbers, we could say: there is an infinite number of numbers of parts into which the lump can be divided. And this is not to say that it can be divided into an infinite number of parts" (2).

---

[1] Throughout, simple parenthetical cites refer to Thomson's "Tasks and Super-Tasks" essay.

It is crucial to grasp the difference. Though there are an infinite number of numbers of parts possible, Thomson does not accept that this means there is some number called 'infinity' using which we can divide the chocolate into 'infinity' parts. "If something is infinitely divisible, and you are to say into how many parts it shall be divided, you have $\aleph_0$ alternatives from which to choose. This is not to say that $\aleph_0$ is one of them" (2).

Let us define more precisely the difference between $\aleph_0$ choices and the choice of $\aleph_0$. Consider the relationship between $\mathbb{N}$ and $\omega$. $\mathbb{N}$ is the set of all natural, or counting, numbers. For us this is the set $\{0, 1, 2...\}$. $\omega$, the first transfinite ordinal, is the first number that comes after all the members of the set $\mathbb{N}$. We can say that $\omega$ is the first number after the infinity of numbers in $\mathbb{N}$. The $\aleph_0$ that Thomson mentions is the *cardinality*, or size as number of elements, of the set $\mathbb{N}$. It is also the cardinality of the set $\omega$, for as we shall see, in Zermelo-Fraenkel set theory with choice (ZFC), the standard construction of numbers is as sets.

Thus Thomson wants us to bear in mind that having a set whose cardinality is $\aleph_0$ does not mean that $\aleph_0$ itself is a member of that set. Here $\aleph_0$ is used somewhat imprecisely, since what Thomson is really talking about is $\omega$, a point he clarifies later on. We can rewrite Thomson's claim as: having all the members of $\mathbb{N}$ as choices does not mean we have the choice $\omega$.

Returning to the chocolate, Thomson blames the imprecision of everyday language for eliding the slip between having $\aleph_0$ many choices and having $\aleph_0$ itself as a choice. "If I say 'It is possible to swim the Channel' I cannot go on to deny that it is conceivable that someone *should have* swum the Channel. But this analogy is only apparent" (3). If one cannot have $\aleph_0$ itself as a choice, then the completion of a supertask remains inconceivable.

It is important to note the tense of his example. Thomson is concerned with whether "someone *should have* swum the Channel," or, whether someone *should have* completed a supertask (3). What is being pointed at is the first moment upon completion of a supertask. Throughout, this point in time is the lynchpin for Thomson.

To illustrate the difference between "an infinity of possibilities" and "the possibility of infinity" Thomson continues dividing his chocolate (3). However, the way in which he does so renders his results mathematically invalid for reasons that will become important in our discussion of infinite series. Recall that Thomson writes, "there is an infinite number of numbers of parts into which the lump can be divided" (2). Further, "each of an infinite number of things can be done, e.g. bisecting, trisecting, etc." (3). Finally, "the operation of halving it *or halving some part of it* can be performed infinitely often" (2, emphasis mine).

What does it mean for an infinite number of numbers of parts to be possible? Thomson began by claiming, "to say that a lump is infinitely divisible is just to say that it can be cut into any number of parts" (2). Strictly speaking there is no need to require cutting into *any* number of parts to guarantee it is infinitely divisible. We could require chocolate to be divided into any number of lumps except three. In this manner we could start with one lump, halve it into two, halve those into four, etc. We would still have infinitely many numbers of parts (since the series can progress without upper bound), but we would have excluded a possible number: three.

While the point might seem pedantic, precision in discussing $\omega$, as well as the limit as one approaches $\omega$, permits our approach to succeed where others have faltered. The reason the series is unbounded on the upper end has to do with $\omega$. Because $\omega$ is a limit ordinal, $\omega - 1 = \omega$. As a first blush of the argument, it is simple to see that if $\omega - 1 \neq \omega$, then there is some number one less than $\omega$ that nonetheless is also greater than all other members of $\mathbb{N}$. However, this contradicts the definition of $\omega$ as the first such number.

Given the three quotations above, Thomson imagines a chocolate-cutter capable of dividing any lump of chocolate with any number of cuts at any step in the process. Our series of chocolate, expressed in total lumps of chocolate, could be: 1, 2, 3, 4, 5, 6… by cutting only one of the previous two sub-lumps. Or it could progress 1, 2, 4…. Or a particularly obstinate chocolate-cutter might refuse to cut at all, yielding 1, 1, 1…. Or, we might choose to cut at some unspecified time X, suddenly transforming from 1 into 2 in a manner analogous to grue: 1, 1, 1… ?, 2.

This malleability of when and how to cut weakens the thought experiment mathematically, for such a series is no longer a mathematical series. A series is a sequence, which is a function on a countable (has cardinality $\aleph_0$) and totally ordered (every element is either greater, smaller, or equal to another) set. A function maps elements from its domain to its co-domain. Crucially, every element in its domain must map to one and only one element in the co-domain.

In our example above one lump of chocolate might map to two, one, or some unspecified probability of either one or two. Two might map to three or four, etc. Hence, this example is not a function and therefore cannot be a series.

**THOMSON'S LAMP INTRODUCED**

The canonical example on the logical impossibility of supertasks is Thomson's eponymous lamp. He describes to the reader "certain reading-lamps" that have a button in the base (5). Pressing the button switches the state of the lamp. If it was off, it is now on, and *vice versa*.

Applying the infinite geometric series $x_n = 1/2^{n-1}$ (which sums to two) where *n* is the term of the sequence, Thomson asks what the final result is if we make one jab for each time interval (1, ½, ¼…). "Suppose now the lamp is off, and I succeed in pressing the button an infinite number of times, perhaps making one jab in one minute, another jab in the next half-minute, and so on…After I have completed the whole infinite sequence of jabs, i.e. at the end of the two minutes, is the lamp on or off?" (5).

Thomson declares the question impossible to solve, since, "I did not ever turn it on without at once turning it off…[and] I never turned it off without at once turning it on. But the lamp must be either on or off. This is a contradiction" (5). Hence the terminal state of the lamp remains a mystery.

Benacerraf successfully challenges this conclusion with the following critique. Demarcating the beginning and final instants of time for the lamp task as $t_0$ and $t_1$ Benacerraf explains, "From this it follows only that there is no time *between $t_0$ and $t_1$* at which the lamp was on and which was not followed by a time *also before $t_1$* at which it was off. Nothing whatever has been said about the lamp *at $t_1$ or later*" (Benacerraf, 768).

He concludes that Thomson has been asking for the impossible by applying conditions about the state of the lamp that are prior to $t_1$ to the lamp at $t_1$. Thus there is no final, $\omega^{th}$ act to determine the state of the lamp. For Benacerraf this does not mean the lamp terminates in a contradictory in-between state, but rather that the contradiction Thomson uncovers is a false contradiction.

Thomson acknowledges the lack of a final, $\omega^{th}$ act in his rebuttal to Benacerraf, writing, "If the successive transitions of *S* [the system] are caused by successive acts of an agent, then for the purposes of this *Gedankenexperiment* we want the agent not to perform an $\omega^{th}$ act; it would only get in the way. But then there is still a last uncaused transition, and it is this that we want to inquire about" (ZP, 134).[2]

From this discussion we take that Benacerraf is right to vitiate Thomson's proof of the paradoxical nature of supertasks on the grounds that the condition of the lamp immediately reversing state only applies to moments in time before $t_1$. However, Thomson is also correct in his rebuttal to note that the central question remains unanswered: *What state is the lamp in at the conclusion of this supertask?* If no such description of the lamp's final state can be given, we might follow Thomson and conclude it inadvisable to speak of supertasks after all.

**RUNNING [0, 1)**

To give a description of the lamp's state at the end of the supertask we must define precisely what moment in time corresponds to the end of the supertask. Considering another example helps clarify the situation.

Imagine a runner who must pass through all the mid-points between the interval *Z* that is [0, 1). He must touch ½, ¾, etc. Since the interval is closed on the left and open on the right, 1 is not part of *Z*, but 0, his starting point, is (10). "Further: suppose someone could have occupied every Z-point without having occupied any point external to Z. Where would he be? Not at any Z-point, for then there would be an unoccupied Z-point to the right. Not, for the same reason, between Z-points. And, ex hypothesi, not at any point external to Z. But these possibilities are exhaustive" (10).

---

[2] Throughout, ZP refers to Thomson's essay rebutting Benacerraf in *Zeno's Paradoxes*.

Benacerraf's response to this paradox is exactly the same as his response to the lamp. Since some time after the completion of the supertask must come (else, the supertask's time to complete is unbounded and therefore it does not complete), conditions about the position of the runner given by the supertask can apply only to positions before time $t_1$.

In his response Thomson later acknowledges he thought of the [0, 1) example "as a kind of joke," since "[t]he resulting situation is simply that [the runner] occupies 1" (ZP, 130). Nonetheless he acknowledges the force of Benacerraf's counter-argument, both with respect to his construction of the lamp and his attempt to analogize the [0, 1) paradox to the lamp (ZP, 130-31). He maintains that there exists "some conceptual difficulty about the idea of a lamp having been turned on and off infinitely often, because, roughly speaking, of the question about the state of the lamp immediately afterwards," and that this concern is independent of the particular way in which he sought to find a contradiction (ZP, 130, 131-32).

There is still something to be gained from Thomson's thought experiment on the interval [0, 1), and that has to do with the status of $\omega$ and $\omega+1$ as ordinals, or numbers that denote position (1st, 2nd, etc.). Thomson notes that the runner who successfully runs an infinite sequence of midpoints from 0 to 1 (and arrives at 1) "is not a sequence of type $\omega$ but a sequence of type $\omega+1$ (last task, no penultimate task), the sequence of the points 0, ½, …

, 1 in Z's closure" (12). A sequence of type ω has no last task, since there is no number immediately 'before' ω. ω+1 does have a last task, the one taking the sequence from ω to ω+1.

Our sympathies lie with Thomson here, for it is precisely the closure of the interval [0, 1) through the addition of the point 1 that is brought about by taking the series from ω to ω+1. This is the same reason why in his rebuttal Thomson points to the fact that the $ω^{th}$ transition remains exactly on the bound. Considering Thomson's Lamp we find an intuitive application of this distinction. The infinite geometric series $x_n$ has only ω many in-order terms. Thus, ω+1, in a sense to be made more precise shortly, refers to the first term that is not actually a member of the geometric series. That is to say, the first moment in time after the completion of the supertask, after the position ω.

To clarify what is meant by ω and ω+1 we need now to pause and consider numbers.

**NUMBERS, TRANSFINITE AND ORDINAL**

As previously stated, ω is the first transfinite ordinal, the first number larger than all members of ℕ. ω+1 is simply the next number. But how shall we define numbers, and how can we define ω? Following von Neumann's standard construction of ordinals,[3] we hold that the ordinal numbers are sets that include all the numbers below them starting from and inclusive of zero.

---

[3] Cf. von Neumann, John. "On the introduction of transfinite numbers." (1923). *From Frege to Gödel: a source book in mathematical logic, 1879-1931*. ed. Heijenoort, Jean. Cambridge: Harvard University Press, 1967.

For example, the number 5 is {0, 1, 2, 3, 4}. Likewise, the number ω must include all the members of ℕ. Note that a number is not present in its set: 5 does not appear in 5. Note also that the cardinality of the set is equivalent to the name we give it (the cardinality of the set 5 is in fact 5), a point we return to in our solution to Thomson's Lamp.

Movement up the chain of natural numbers is given by the successor function *S(a) = a* ∪ *{a}*, with 0 defined as the empty set, ∅. Thus 1 is ∅ ∪ {∅}, which yields {∅}, the set containing the empty set. Alternatively we may write 1 as {0}. To get 2 we reapply the successor operation, yielding S(1) = 1 ∪ {1}, which gives {0, 1}, etc. However, ω presents a different case.

ω is a special ordinal because it is a limit ordinal.

> A limit ordinal is an ordinal number that is not a successor ordinal. Specifically, an ordinal W is a limit ordinal if and only if there exists some ordinal x < W, and for any x < W there exists another ordinal y such that x < y < W.[4]

We will comment on this extraordinary definition shortly, but first we must ask, how can we generate a limit ordinal? A limit ordinal cannot be given by the successor operation since every ordinal given in such a manner can produce another ordinal greater than it by one more application of the successor operation.

---

[4] An alternative but equivalent definition is offered on p. 20 of Jech, Thomas J., *Set Theory*. Third Millennium Edition. Springer, 2003. We offer the above formation for clarity of argument.

The operation required to produce a limit ordinal is the *supremum*.[5] A limit ordinal is the supremum of all the ordinals below it, which is given by taking their union. In the case of a normal, finite number we can see that the supremum exists conventionally speaking, as the maximum.

For the limit ordinal ω there is no ordinal immediately prior to it, by definition. It is against this infinite succession of ordinals that one must take the union, and in so doing produce a supremum where no maximum exists, thereby generating ω. Once generated, ω is free to appear as an element in the set of a larger ordinal via the successor operation, thereby paving the way for ω+1, ω+2, etc.

This generation of a supremum without a maximum may seem scandalous. Mathematically speaking, it stands on solid ground. Present in the ZFC axioms is not only an Axiom of Infinity (an infinite set exists), but also an Axiom of Union (defining the union operation). ω is what happens when infinite successor operations are possible and the union of that infinite sequence is taken. It is a supremum of a set with no maximum, made possible by an Axiom of Union that is every bit as powerful as the Axiom of Infinity.

---

[5] The meaning of the word supremum varies with context. In some contexts ℕ has no supremum, e.g. when it is embedded in ℝ. In ordinal theory every set of ordinals has a supremum.

Returning to our runner, what are we to understand when Thomson writes that the runner at position 1 "is not a sequence of type ω but a sequence of type ω+1 (last task, no penultimate task), the sequence of the points 0, ½, … , 1 in Z's closure" (12)?

Were the task to remain an order-type of ω, it would still involve members of ℕ, which means that ω itself would never be reached. This much follows readily from our construction of ordinals as containing the whole numbers below them, starting from and inclusive of zero. But what does it mean to generate ω, that is, to *reach* ω only through a task of order-type ω+1?

It means that the supremum alone is insufficient, but instead a supremum plus an additional successor operation, *S(ω)*, is required. Seen in this light it becomes clear that for all supertasks, questions of their completeness must necessarily be about order-types ω+1.

With the distinction between ω and ω+1 clear, we can now tackle the question of how to evaluate a supertask's completion more precisely. We know the evaluation must take place at point ω+1, but is it an evaluation of the function representing the supertask, or an evaluation of the state of the world? If it is the former, the function is undefined at ω+1 (the supertask is finished), which is exactly why Benacerraf's critique is trenchant. If it is the world, then we must ask what is meant precisely by ω+1 in the context of the world.

Is it not instead *world + 2 minutes*[6] (assuming the ω tasks are accomplished, along with the final transition of the system at moment ω, within two minutes)? The following quote from Thomson's rebuttal to Benacerraf is worth reproducing in full.

> "In general, the idea of an ω-task arises from consideration of a bounded ω-sequence of points on the real line. If the sequence is not bounded there is not even the semblance of a likelihood that the task can be completed; if it is bounded, the sequence will have a least upper bound and there will be some question to ask about the state of the world or about what happens at some time corresponding to that bound" (Thomson, *ZP*, 134).

It is worth recalling that the number of *terms* in the time-sequence remained of order-type ω, with cardinality $\aleph_0$. Consequently, to speak of an order type ω+1 requires us to speak of the smallest possible first instant of time within which the supertask is complete, which is when our analysis shifts focus from the function representing the supertask to the state of the world.

**THOMSON'S LAMP SOLVED**

With this preparation in place we ask the following question. Can we establish that there exists a supertask that is logically possible? If yes, then we have solved Thomson's initial claim that supertasks are inherently paradoxical. To do so we must answer the nexus question: *What state is the lamp in after two minutes of infinite button presses have elapsed?* We maintain the following original solution constitutes the first known solution to Thomson's Lamp.

---

[6] Technically, as we will see, it is *world + 2 minutes + ψ* where ψ is the limit as *n* tends to infinity of $x_n = 1/n$. In other words, it is the first instant in time following the completion of the supertask, and is in some sense external to the completion of the supertask itself (which is why the sequence defining the supertask is undefined at this point). Here we see it expressed in ψ, using the language of infinitesimal limits.

The lamp maintains its original state. As Thomson notes, an even number of switches will maintain the lamp's state while an odd number will change it (5). Thus the only question that requires answering is, "Is infinity even?" It is, when we use the set-theoretic definition of even to mean capable of subdivision into two disjoint subsets of equal cardinality. As we will show, this set-theoretic definition of even is equivalent to the numerical, "divisible by 2" definition to which we are accustomed.

It is trivial to show that a countably infinite set can be divided into two disjoint subsets of equal cardinality. Consider $\mathbb{N}$. From the definitions for even and odd we know that if a given integer $j$ is even, $j+1$ is odd. From the definition of $\mathbb{N}$ we know that for all $j \in \mathbb{N}$, $j+1$ exists. Accordingly we can subdivide $\mathbb{N}$ into two disjoint subsets of evens and odds respectively, both of which have the same cardinality.[7]

One might object to our usage of the set-theoretic definition of even since Thomson was asking a question about the *number* of tasks performed, and of what *number* (1 or 0, on or off) would represent the lamp's final state. However, the two definitions are equivalent,

---

[7] They happen to also have the same cardinality as the original set $\mathbb{N}$, since all three sets are countably infinite. This quality is at the root of Galileo's perfect squares paradox.

once the consequences of a set-theoretic construction of numbers is understood more fully.

We traditionally define even and odd as follows.

    Eqn 1.1    An even integer n = 2k where *k* is some other integer

    Eqn 1.2    An odd integer m = 2k + 1 where *k* is some other integer

We can rewrite these definitions in terms of our earlier set-theoretic definition of number. Let us define a natural number *r* as a set *s* that contains all natural numbers that precede *r*, starting from and inclusive of 0. So again we have the integer 5 defined as the set {0, 1, 2, 3, 4}. Note that the cardinality of *s* is equivalent to *r*. Let us use the set-theoretic definition of odd and even to define the oddness or evenness of set *s*. If set *s* can be divided into two disjoint sets (whose union contains all its elements) that are equinumerous (whose cardinalities are equivalent), then we say set *s* is even. If not, set *s* is odd.

Let us denote the sets *u* and *v* as the two disjoint subsets into which we divide *s*. Let us denote the cardinalities of *u* and *v* as *w* and *x* respectively. If and only if *w* = *x*, can we say that the set *s* has successfully been divided into two disjoint subsets of equinumerality. However, if *w* = *x*, then we can express *r*, the cardinality of set *s*, as 2 * *w*. If not, we express *r* as 2 * *w* + 1, since at most the count can be off by one (since we have been choosing from among two subsets, and thus if we had two extra items left in our parent set, then each subset would have received one, etc.). Thus we have arrived at our so-called 'numerical' definition of even from the set-theoretic.

We can see that the two disjoint subsets into which we divide the button presses of the Thomson Lamp can be none other than the sets of 'switch to on' and 'switch to off,' just as they appeared in Thomson's use of the Grandi sequence (6). [8] What is novel is the knowledge that the termination of the supertask depends, in this case, on the initial configuration. If the lamp was on, it remains on at the conclusion. Else, it remains off.

One might object that even though the cardinalities of the two subsets are equivalent, both are $\aleph_0$, which is the same cardinality as our original set $\omega$. Hence, it might be impossible to determine which of our $\aleph_0$ subsets terminates 'after' the other, voiding our application of evenness to the proof.

Such worries are easily put to rest once we view the $\aleph_0$ cardinality of the parity proof's two disjoint subsets as a restatement of *$2\omega = \omega$*. That $2\omega = \omega$ is clearly understood once order-types are applied rigorously. Since $\omega$ is an ordinal we know it is specifying a positioning, or sequencing of elements. To say that $2\omega = \omega$ is simply to say that $\omega$ pairs in sequence possess the same order-type as $\omega$ singletons in sequence.

---

[8] The sequence is $1 - 1 + 1 - 1 \ldots$. Thomson assigns the states '0' and '1' to 'lamp off' and 'lamp on,' and asks what the sum of Grandi's sequence can tell us about the lamp's final state. In our opinion the question is misguided, as the sum of this divergent series is given as ½, which is not a reason for supertasks being internally self-contradictory, as Thomson supposes, but instead is due to the particular definition of a Cesàro sum, whereby a second converging series is constructed by the partial means in order to find a consistent value to assign as the sum for certain divergent sequences.

At first this explanation does not seem to extricate us from our quandary, since the pairs are obviously (switch, switch) pairs of button presses (Thomson's claim of never once turning the lamp on without turning it off and *vice versa*). However, we can use our initial knowledge about the lamp's state at the beginning of the experiment to fix the state of the lamp upon completion. Since the ω pairs of (switch, switch) will take place regardless of what state the lamp is in at the beginning, and since we know that an even number of state transformations preserves state, we can see that the initial state of the lamp can be removed without affecting the even number of state transformations that take place. And in fact, it is removed, since the initial state does not count as a button press. Hence, the initial state remains the final state as previously described.[9]

Finally, an inquisitive reader might inquire if $3\omega = \omega$, thereby demonstrating an arbitrary decision to subdivide into two disjoint subsets instead of three.[10] Perhaps this argument points to slippage between the set-theoretic and numerical definitions of parity, which earlier were claimed to be equivalent. No such slippage is evident. That a number (set) may also be divided into thirds (or fourths, or…) does not mean it cannot also be even.

---

[9] For another way of thinking through this proof, we might consider what was at stake when Cesàro sums became the accepted way of fixing the sum of certain infinite divergent series. The need for a way to fix the sum was precisely because alternative methods, all of which seemed reasonable, gave inconsistent results. If one took Grandi's sum as 1 + (-1 + 1) + (-1 + 1)… then the answer was 1. If one grouped the terms as (1 - 1 ) + (1 - 1)… then the answer was 0. We can view the initial state of the lamp as an instruction on how to group the terms.

[10] For example, the three subsets $\{n \in \mathbb{N} \mid n \bmod 3 = 0\}$, $\{n \in \mathbb{N} \mid n \bmod 3 = 1\}$, $\{n \in \mathbb{N} \mid n \bmod 3 = 2\}$. In this manner arbitrarily many disjoint subsets of equal cardinality may be constructed from $\mathbb{N}$. Thanks to [REDACTED] for this formulation, and much kind counsel besides.

**SUPERTASK COMPLEXITY**

We can now put to rest Thomson's anxieties about having a "method for deciding *what* is done when a super-task is done" (6). However, the generalizability of our results is an open question. "One difficulty, then, about the question of whether ω-tasks can be completed is that there are different kinds of them, and there is no reason to think that in regard to completability they stand or fall together" (ZP, 136). Nonetheless, it is a significant result to be able to declare that *there exists* a supertask for which a logical rendition of the state of the world following its completion may be given. That this supertask happens to be the *ur*-supertask is a happy coincidence.

Regarding the generalizability of our results, we now prove the following theorem.

> Theorem 1: Any $\omega_0$-task describable by a state transition system with a period of length less than $\omega_0$ reachable by a path of length less than $\omega_0$ has a solution completely determined by its starting state.

$\omega_0$-task refers to supertask in the sense we have been using it in this paper. Thomson will often abbreviate this as "ω-task," a convention from which we briefly depart for technical precision. Period refers to a cycle (e.g. ending where you began) in a state transition system. The length of a period refers to the number of transitions made to complete one cycle.

We now define a mathematical object, inspired by automata theory, which we call a "state transition system" (abbreviated STS).[11] An STS consists of:

- Set of states
- Set of starting states
- Transition function
- Runtime
- Starting state selection

Optionally, a diagram like the one below may accompany the STS. The set of states refers to all individual states the system can occupy (the "state space"). The set of starting states is a subset of the state space that identifies valid initial states for the system. The transition function describes the next state in terms of the current state, and is thus a function from the set of states to the set of states.

The requirement that the transition function be a function is slightly stricter than simply equating it to a set of edges and the STS to a weakly connected graph.[12] Requiring a function necessitates only one outgoing edge from each node (state). However, the requirement for a transition function does not exclude disconnected cycles,[13] provided each cycle is accessible from a valid starting state (else, it is not a set of nodes the system can occupy, and thus is not a member of the system's state space).

---

[11] The interested reader might pursue Deterministic Finite State Machines for more. The differences between DFSM and STS have largely to do with the use of DFSM to determine acceptable words in a language, whereas, STS have been explicitly constructed to describe supertasks.

[12] A weakly connected graph is a directed graph (edges have arrows) in which, if one ignores the direction of the edges, there is a path from any node to any other node.

[13] A disconnected cycle refers to a cycle that, once entered, does not permit exit outside the cycle. Such cycles are possible due to the directed nature of the edges.

While not strictly part of the STS, the runtime defines the length of the operation the STS will be used to evaluate in terms of number of transitions. For supertasks as described in this paper the runtime is $\omega_0$. Similarly, the selected starting state is not necessary for the definition of an STS, but may be necessary for the description of a particular run of an STS.

For the Thomson Lamp the set of states is {off, on}. The set of starting states is likewise {off, on}. The transition function is a simple negation: {off → on, on → off}. The runtime is $\omega_0$. The selected starting state is {off}.

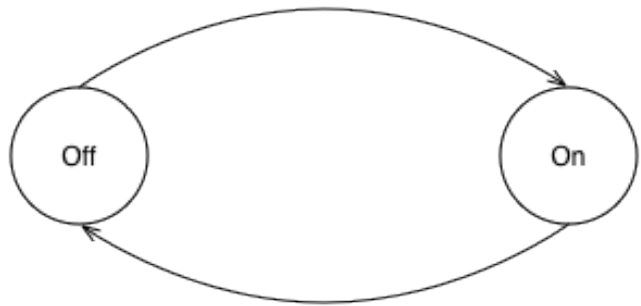

*Figure 1. Thomson Lamp STS*

The proof of Theorem 1 above follows naturally from our definitions concerning STS's and periods. Recalling the discussion of the parity of $\omega_0$ from the previous section, given that the length of the period $k$ is less than $\omega_0$, we know we can form $\omega_0$ many *k-tuples*.[14]

Hence, after the completion of $\omega_0$ many transitions we have actually completed $\omega_0$ many traversals of the cycle described by the period and the final state is the same as the initial state.

The theorem holds even in cases where there exist starting states from which no cycle is accessible, since it claims only that given a cycle (with finite period length) reachable by *at least* one initial state, the end result for this starting state is determined entirely by the starting state. Note that this holds even when the initial state is not a member of the cycle, by an extension of the argument that $\omega_0 - 1 = \omega_0$ for all finite replacements of 1.

> Theorem 2: The solution in all cases described by Theorem 1 is always the entry node of the accessed cycle.

Where 'entry node' refers to the first node encountered that is part of the cycle, and 'accessed cycle' refers to the cycle the run traverses.

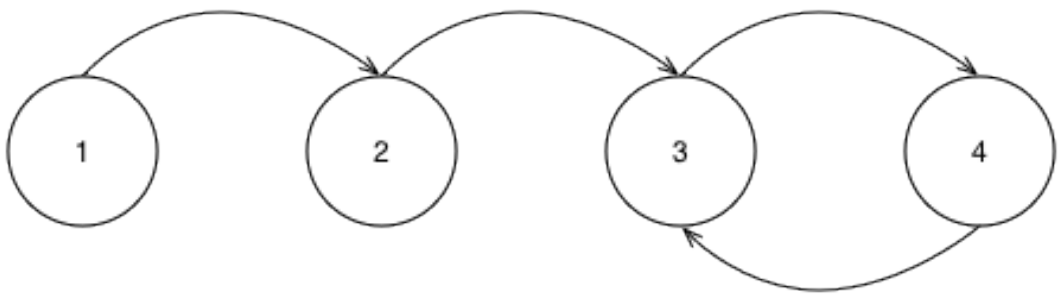

*Figure 2. A simple STS with a cycle*

---

[14] In other words, if $k = 2$, we can form $\omega_0$ many pairs in sequence. If $k = 3$, $\omega_0$ many triples in sequence, etc.

For example, in *Figure 2*, the solution for start states 1, 2, and 3 is 3, while the solution for start state 4 is 4.

A host of questions presents itself about the capacity of STS's to accurately describe the behavior of various supertasks. Theorem 1 seems to generalize for all countably infinite supertasks, e.g. all $\omega$-tasks where $\omega$ is some transfinite ordinal $< \omega_1$. Note that Theorem 2 holds only if the order-type of the supertask is a limit ordinal.[15] Precisely defining the ordinal relations between runtime, period length, and access-path length (e.g. length of path to access the cycle from a given starting state) remains to be done.[16] Similarly, questions arise about the capacity of an STS to represent supertasks with transfinite state spaces (state sets with cardinality $\geq \aleph_0$).

Can STS's be extended to describe supertasks with non-deterministic transition functions, or supertasks whose runs include non-deterministic selection of starting states? How to describe a supertask with a multi-vergent transition function, e.g. one with multiple outgoing edges from a state, and where a given state is a composition of several state nodes (perhaps with a transition function on the power set?). Can the transition function rely on more history than just the previous state?

---

[15] E.g. it holds for $\omega_0$ and $\omega_0 2$, but not for $\omega_0+2$ nor $\omega_0+1$. The solution remains simple to compute with knowledge of the 'remainder' $q$ after $\omega_0$: simply take $q$ more transitions from the entry node of the accessed cycle.

[16] In our example access-path and period length require being on the lower side of a limit ordinal where the runtime is that limit ordinal. This relation may generalize to limit ordinal runtimes $> \omega_0$, thereby enabling access-path and period lengths that are still infinite, but proportionally smaller, limit ordinals.

Given responses to these questions it may be possible to define a hierarchy and typology of STS's, providing a complexity classification for supertasks. It may even be the case that certain types of supertasks foreclose the knowability of their solutions, as may be the case with Thomson's π-parity machines. In this example Thomson proposes one machine elaborating the digits of π, and another machine outputting 0 or 1 based on the oddness or evenness of the corresponding digit.

π's status as an irrational number guarantees the lack of any period with length $< \omega_0$, though this leaves open an intriguing question concerning STS's with runtimes of ω2 or greater. Even more problematic, we lack a meaningful transition function based on the set of available states the decimal expansion of π can occupy, {0, 1…9}, as current methods for calculating π rely on iterative approaches or approaches based on the particular digit of π one wishes to calculate.[17]

Further study on this question will no doubt make use of the rich resources available in the study of computational complexity and recursion theory such as Turing degrees and Turing jumps, Turing machines and oracles, and Markov processes. Indeed, some work in this direction has already begun.[18] With a solution to Thomson's Lamp in place, the way is now clear for a logically defensible discussion of hypercomputation freed from the trappings of pure fancy.

---

[17] Cf. Bailey, D.H., Borwein, P.B., and Plouffe, S. "A New Formula for Picking off Pieces of Pi," Science News, v 148, p 279 (Oct 28, 1995).
[18] See Welch, P.D.. "Turing Unbound: on the extent of computation in Malament-Hogarth spacetimes." British Journal for the Philosophy of Science 59.4 (Dec. 2008): 659-674.

## ORDER (POSITION) VS. SIZE (DISTANCE)

We retreat now from the tantalizing opportunities presented in the model of the state transition system to consider more closely the philosophical ramifications of transfinite ordinals, as well as their bearing on both Thomson and Benacerraf's original formulations.

The transfinite ordinals do exactly what their name suggests: provide us a way to talk about position that traverses infinity. We consider the formal definition of limit ordinal given earlier.

> A limit ordinal is an ordinal number that is not a successor ordinal. Specifically, an ordinal W is a limit ordinal if and only if there exists some ordinal $x < W$, and for any $x < W$ there exists another ordinal y such that $x < y < W$.

$\omega$ is a limit ordinal precisely because there is an ordinal smaller than it, and for all ordinals smaller than it there is an ordinal larger than that smaller one that *is still smaller than* $\omega$. Herein lies the beauty of $\omega$. There is no largest member of $\mathbb{N}$ and yet $\omega$ exists. In fact, we define $\omega$ in precisely this fashion.

Benacerraf is right to point out that the confusion surrounding Thomson's paradox may be due in no small part to the insufficiency of language colliding from two different domains, here the notion of a 'task' with the notion of the 'infinite' (Benacerraf, 782). If we rely on the intuitive notion of tasks as discrete operations, ω never arrives since by definition there is always one more task to complete.

Nonetheless, by bracketing this infinite movement of ω within a definition, we may then talk about and even beyond ω, which is the beauty of the mathematical form. It is a logical and powerful consequence of the Axioms of Union and Infinity taken together.

There is still something to be said vis-à-vis the suspicion with which one might view the definition of ω, given the lack of a largest member of ℕ. This anxiety marks something of significance for the philosophical consequence of Thomson's Lamp, and transfinite ordinals more particularly. Namely, they inaugurate a concept of position or order decoupled from size or distance. Intuitively speaking, we conflate order with size regularly. The first place has the smallest distance from zero. The second the second smallest, and so on.

What must be stressed is the intimate relation between distance from zero and size. Recall our nested definition of numbers. Recall as well that the cardinality of such a set is equivalent to the number that set is. Distance from zero is enumerated via the successor function, which successively adds one element to the set. This enlargement by one corresponds to a positional increase by one as well, moving us from an ordinal to its successor ordinal. The correspondence between increase in size and increase in position has been necessary, logical, and irreversible.

Only in the limit ordinal does the relationship between the cardinality of the set in question and its place as an ordinal become unmoored. By definition the number of ordinals less than $\omega$ is infinite. Specifically, it is countably infinite. To say a set is countably infinite is to say that it can exist in bijection (one to one correspondence) with $\mathbb{N}$. Following Cantor, whose work birthed the study of the transfinites, we say all these sets have cardinality $\aleph_0$. Even the sets $\omega+1$, or $\omega 2$ (the second limit ordinal) have cardinality $\aleph_0$ (the first transfinite with cardinality greater than $\aleph_0$ is $\omega_1$).

Strictly speaking, then, we cannot say that the number of numbers in the set $\omega$ is greater or less than the number of numbers in the set $\omega+1$. Yet, we *can* say that $\omega+1$ occupies a different position from $\omega$. This is the first time in our discussion of numbers' positions that two numbers with equal cardinality have been identified as occupying different positions. It is likewise the first time that size and order are no longer co-extensive, hence our earlier criticism of Thomson's slippage between $\aleph_0$ and $\omega$ in our discussion of chocolate.

**THE METRICAL AND THE ORDINAL**

What are we to make of this apparently paradoxical decoupling of position and size? In reality, the alliance between size and position belies a much more profound alignment, active but unnoticed, in the analyses of Thomson and Benacerraf. Both pass over a remarkable relation between two very different kinds of limits employed: the metrical, and the ordinal. Thomson in his rejection of infinite sums as proof of paradox in supertasks, and Benacerraf in his equivocation between the approach to an upper-bound on the reals and the completion of the supertask.

These two limits will become clearer as we consider summations of infinite series, which Thomson takes to be self-contradictory (8-9). First, some definitions. A metric, or distance function, provides a distance between elements of a set. A metric space is a set with a metric. The Euclidean space of dimension one is the real number line.

Consider the definition of the metrical limit.

> *lim* is the limit of a sequence $x_n$ if, for all $\varepsilon \in \mathbb{R}$ where $\varepsilon > 0$, there exists some $n \in \mathbb{N}$ such that for all $m \in \mathbb{N}$ where $m > n$,
>
> $|x_m - lim| < \varepsilon$.[19]

What must be emphasized here is the crucial difference between the limit at work in the sum of an infinite series and the limit at work in the first limit ordinal, $\omega$. We see two infinites, moving in opposite directions. First the infinity of the increase in the input term, the $n$ in $x_n$. For this infinity only a cardinality of $\aleph_0$ is necessary, so the set $\mathbb{N}$ suffices.

---

[19] An equivalent definition is offered on p. 38 of Courant, Richard, and E.J. McShane. *Differential and Integral Calculus*. 2nd ed. Vol. 1. London and Glasgow: Blackie & Son Limited, 1937. We offer the above formation for clarity of argument.

We have as second infinity the closeness measure ε, which may be made arbitrarily small provided it remains greater than zero. What our definition says is that no matter how small we make ε, there exists some natural number beyond which the distance between outputs of the function and our limit is less than ε. Thus, no matter how close to the limit we wish to force our output, we are capable of doing so.

This second infinity requires an infinity of cardinality $2^{\aleph_0}$, that of the set of reals $\mathbb{R}$. In each of Thomson's supertask examples (chocolate, [0, 1), the lamp) there has been a tacit acknowledgement of domain $\mathbb{N}$ for the function representing the supertask. This domain has functioned indexically, providing the current position in the execution of the supertask. The co-domain for the output of this function has been different in each case: $\mathbb{N}$ (total lumps of chocolate), $\mathbb{R}$ (position on the interval [0, 1)), or {0, 1} (lamp off or on) respectively.

However, in the latter two supertasks there has also been a *second* function, that of the timekeeper. This function maps the index of the supertask's operation to the reals via the geometric sequence $x_n = 1/2^{n-1}$. This sequence provides the successively decreasing intervals of time in our two minute march to infinity.[20]

Any supertask requires a bounded period of time for its execution. Thus, there will always be some infinite series on the reals to provide a decreasing time per execution. It is because the set $\mathbb{R}$ of our metric space, to which our distance function applies, is infinite with cardinality $2^{\aleph_0}$ that we can produce arbitrarily small distances with our closeness measure $\varepsilon$. These distances in turn permit an infinite timekeeper series to accelerate our supertask to completion. A cardinality $\leq \aleph_0$ would not provide for arbitrarily infinitesimal intervals to be cut.

---

[20] In the [0, 1) example this function also gives the runner's position, so the supertask function and timekeeper function are mathematically identical. It should also be noted that strictly speaking, the chocolate division was not a supertask as its operation was not finitely bounded by time.

Hence the limit in the timekeeper sense is a measure that *only* makes sense with distance (the metric); whereas, the limit of $\omega$ (which gives position in the supertask sequence) forces us to decouple position (order) from size (distance). So when Thomson asks us to consider the sum of an infinite series as a limit,[21] and uses that to speak of a set of $\omega$ tasks, he appears to be equivocating between two different infinities.

In point of fact this equivocation is likely the cause for Benacerraf's critique, which spends most of its time concerned with the approach to an open upper bound on a segment of the reals (e.g. the vanishing genie).[22] Can we find a way to resolve the tension between metrical and ordinal limits for supertasks?

---

[21] See p. 7-9 of "Tasks and Super-Tasks" where Thomson discusses Watling's arguments about summing convergent infinite series. In particular p. 9 where Thomson concludes that insufficiencies in likening a convergent limit to an actually terminating sequence of sums has some bearing on the logical possibility of supertasks.

[22] The genie's size halves with every advance in the series $x_n = 1/2^{n-1}$ (thus vanishing at the two minute mark) and beautifully illustrates Benacerraf's argument that conditions prior to $t_1$ cannot be used to evaluate the system at $t_1$ or later. As Thomson notes, Benacerraf fails to provide definitive means for evaluating the state of the system at $t_1$ or later.

We maintain that the ordinal and metrical limits are directly related. It is precisely the inexhaustibility of $\mathbb{N}$ (necessary for ω) that allows us to state definitively that there is some sufficiently large $n \in \mathbb{N}$ beyond which our function's output satisfies all possible closeness measures to the metrical limit. In distinction to the "illusion…that one might reach the sum…by actually adding together all the terms of the infinite sequence," what the definition of a function's limit for metric spaces shows us is that the 'sum' here is understood in the only way possible: in relation to a function's output as its input increases without bound (9). What is up for discussion is the behavior of a function under variation, not a finite sum and its numerical terms, a point to which we return in our discussion of Deleuze.

While this explanation compels a reconsideration of Thomson's equivocation between different infinities, have we really answered the most basic objection to the summation of an infinite series: that it could never take place? Should we fail here, must we not also refuse to admit the possibility of any supertask whatsoever (as Thomson concludes)?

Consider what happens if we set the closeness measure to zero. The only argument capable of satisfying that closeness measure is to evaluate the function at $x_\omega$, the first position after the set of natural numbers. We see that the only number capable of reducing distance between series and sum (metrical limit) to zero is precisely the first number that decouples distance from position (limit ordinal). Paradoxically (but only to the non-mathematician), the number that could guarantee distance travelled, as in the Thomson experiment of walking the midpoints from [0, 1), is the number that requires us to let go of distance (size).[23]

**IDENTIFYING FREGE**

What shall we make of the equality established with the reduction of our closeness measure, ε, to 0?

$$\text{Eqn 2} \qquad x_\omega = \lim_{n \to \infty} x_n$$

On the one hand, we appear to have reached ω by walking patiently along $\mathbb{N}$. On the other, we seem to have approached ω by shrinking to zero our closeness measure ε along the reals. Mathematically the confusion is absent. We did not approach ω by shrinking ε, but instead approached $x_\omega$, the $\omega^{th}$ term in the series $x_n$. It is the difference (a real quantity) of this value with the limit of $x_n$ that has shrunk to zero.

Nonetheless the equality above contains a measure of interest for us, insofar as it collapses or otherwise brings into momentary contact the metrical and ordinal limits. How shall we theorize this contact? And how shall we theorize this equation? Let us start with the problem of the equality itself, and with that duly considered proceed to examine the point of contact.

---

[23] For more see "compactification." There are interesting methods of compactification, e.g. bending the real number line into a circle with one point joining +/- ∞, ripe for analysis.

Of immediate interest to us is the equals sign. It is this sign that binds the equation together, providing its motive force and bringing about contact between the two limits. To consider its implications we look to Frege's *Begriffsschrift*. Inspired by Leibniz's call for "a universal characteristic…a *calculus philosophicus* or *ratiocinator*," Frege aims to construct a "pure logic…disregarding the particular characteristics of objects, [that] depends solely on those laws upon which all knowledge rests" (*B*, 6, 5). [24]

Frege begins by defining his symbols. Of particular interest to us is §8 "Identity of content," where the sign '=' is discussed (*B*, 20). It appears as though this section simply asserts a transparent meaning for the '=' sign, which Frege labels the "identity of content" sign (*B*, 20). Namely, that the figures on the left and right share the same content. However, the field opened by the identity sign is more complex than it seems.

Professor Bar-Elli's excellent reading of §8 opens the way for our analysis. There exists a tripled field between name, sign and content that the identity sign arbitrates. Frege is direct, opening the section with an unambiguous sentence where he shifts from speaking of signs to names. "Identity of content differs from conditionality and negation [two symbols previously introduced] in that it applies to names and not to contents" (*B*, 20).

Bar-Elli describes the magnitude of this terminological shift, writing that "[t]hroughout this section, except for the last sentence, Frege speaks of identity consistently in terms of names, i.e. signs endowed with modes of determining their contents, whereas in the rest of the book he talks, where identities are concerned, simply of signs" (Bar-Elli, 357). A sign "just denotes its content; this exhausts its meaning. A name, in contrast, includes a mode of determination (Bestimmungsweise) of its content" (Bar-Elli, 357).

---

[24] Throughout, citations marked 'B' refer to *Begriffsschrift*.

In typical usage we have a sign stand in for its content. "The correlation between a sign and its content is an arbitrary convention or stipulation" whereby a sign expressly links its content to the text and is exhausted in this operation (Bar-Elli, 358). "The semantics of names, in contrast, is 'thick': a name does not only denote its content, but includes and expresses a way its content is determined. This, Frege emphasizes, is an objective feature that pertains to the 'essence of things' (Wesen der Sache), to use his terms" (Bar-Elli, 358).

A name must carry its own history with it. This history is in fact an objective and essential feature of the name and its content. It forms a rich, 'thick' linkage between name and content that is "not arbitrary or conventional" (Bar-Elli, 358). Frege himself is quite explicit on this matter. "At first we have the impression that what we are dealing with pertains merely to the *expression* and *not to the thought*, that we do not need different signs at all for the same content and hence no sign whatsoever for identity of content. To show that this is an empty illusion I take the following example from geometry…" (*B*, 21).

Frege goes on to demonstrate how two different points A and B on a circle are actually the same, once it is understood that the differing specifications given for each point yield the same position on the circle's circumference. However, absent the proof that the two points are in fact identical, no such statement could be made. Thus it is not a simple matter of a redundant sign for identical contents, but instead a matter of necessity that an identity of content sign demarcates a relation of equivalence between two contestable micro-genealogies, the thick descriptions of the name-content relations embedded in each of the names on either side of the '=' sign.

In a remarkable passage Frege writes,

> "Whereas in other contexts signs are merely representatives of their content, so that every combination into which they enter expresses only a relation between their respective contents, they suddenly display their own selves when they are combined by means of the sign for identity of content; for it expresses the circumstance that two names have the same content. Hence the introduction of a sign for identity of content necessarily produces a bifurcation in the meaning of all signs: they stand at times for their content, at times for themselves" (*B*, 20-21).

The distinction between sign and name is no doubt the chief innovation of §8, as Bar-Elli notes. Far from a universal logic whose application depends not at all on the particulars of the objects before it, this logical sign reveals the others as very much *alive, contestable,* and *inseparable from their histories*. The sign responsible for arbitrating identity of content produces an inherent bifurcation in all signs. Signs come to "stand at times for their content, at times for themselves" (*B*, 21).

Returning to Equation 2, we now note that the identity sign effects a doubling in each of the terms surrounding it. Reading the equation requires entering each sign, converging on the self-same content the two share, and then rebounding along different paths to return to their names, tracing an echo along each micro-history. The import of the equation is not simply that the two signs express the same content, nor less still that the two names are different ways of expressing the same content. Rather, the significance is that the two ways of determining expressed by each name, while distinct, have converged upon the same concept. It is precisely in the operation of the identity symbol itself that we find this yoking of difference, an expression that in this paper cements the resolution of a paradox.

**FOLDING DELEUZE**

Within the dual compression and expansion of the content and its two names' histories we find the point of contact between the metrical and ordinal limits. We turn now to investigate the nature of this contact.

Recall that in the chocolate task Thomson argues $\omega$ (though he misidentifies it as $\aleph_0$) is never reached. Though the chocolate-cutter continues dividing without rest, we never reach an 'infinity' of chocolate pieces. We simply experience an infinity of numbers of pieces into which the chocolate can be divided.

This argument is analogous to saying that $\omega+1$ never arrives. Since we present $\omega+1$ as the first instant in time following the completion of the supertask, it necessarily falls outside of the function defined over the duration of the supertask, e.g., there is no $x_{\omega+1}$. Thus it is the presence of a bounded interval on the reals (i.e., the two minutes of button presses) that elicits $\omega$ and brings forth the completion of the supertask.

Speaking precisely, it is the infinition[25] in the approach to the upper bound on the reals driven by decreasing towards zero the closeness factor $\varepsilon$ (metrical limit) that joins with the infinition in procession of natural numbers towards the $\omega^{th}$ transition of the system representing the state of the task (ordinal limit). There is a special structure at work here, which following Deleuze we identify as the fold. The fold occurs between the doubled use of the ordinal limit to drive not only the evaluation of the metrical limit (the closure of the finite time series on the reals occurs at argument number $\omega$), but also *the state of the supertask system itself* (i.e. the resting position within the STS defined above is the result of transition $\omega$).

It is no mistake that we find the shape of our contact, first illuminated by Bar-Elli's reading of Frege's identity sign for content, in Deleuze's *The Fold* (subtitled appropriately "Leibniz and the Baroque"). The fold is precisely the shape of the structure that links the dual separation of each sign from its content and itself on the one hand to the convergent commonality of their distinct micro-histories on the other.

---

[25] Borrowed from Levinas, this term refers to the process of producing infinities. We may transliterate it mathematically to emphasize the motive quality of infinite successor operations, each producing the argument for the next. See p. 25-26 of Levinas, Emmanuel. *Totality and Infinity: An Essay on Exteriority*. The Hague: M. Nijhoff ;, 1979.

As we saw previously, this contact is brought about by the consideration of these two limits, metrical and ordinal. We mentioned that the ordinal is doubled, but in point of fact so is the metrical. For it is so not only in the decrease towards zero of ε, but also in the increase towards the upper-bound of the real time interval. These two sites correspond, in what is already another fold, to the ordinal's role in the evaluation of the closure of the time interval and the ordinal's role in presenting to completion the state of the supertask system, respectively.[26]

What has been instrumental from the outset (and something both Thomson and Benacerraf were acutely aware of) is the consideration of infinite sums, and thus limits. "The definition of Baroque mathematics is born with Leibniz. The object of the discipline is a 'new affection' of variable sizes, which is variation itself" (*F*, 17).[27] The key of the new differential calculus is that variability itself is made to vary.

> "To be sure, in a fractional number or even in an algebraic formula, variability is not considered as such, since each of the terms has or must have a particular value. The same no longer holds either for the irrational number and corresponding serial calculus, or for the differential quotient and differential calculus, in which variation becomes presently infinite" (*F*, 17).

---

[26] The associations crisscross precisely because the value of the metrical limit is the closure of the time interval on the reals (e.g. the 2 minute mark in the lamp case).
[27] Throughout, '*F*' in citation refers to *The Fold*.

Though he was unaware of it, Thomson's rejoinder to Benacerraf in *Zeno's Paradoxes* already contained the key to solving his original supertask: "[i]n general, the idea of an ω-task arises from consideration of a bounded ω-sequence of points on the real line" (ZP, 134). Deleuze solves the puzzle for us. It is in the transition from discontinuous variation to continuously variable variation, in other words the advent of calculus of the infinitesimals, that "exposes" the "straight line of rational points…as a false infinity, a simple undefinite that includes an infinity of lacunae" (*F*, 17).[28]

"In short, there will always be an inflection that makes a fold from variation, and that brings the fold or the variation to infinity" (*F*, 18). Without the bounded segment on the reals there was insufficient infinition to bring about the appearance of ω, thinkable (as we know from our set theoretic study of the ordinals) only with an imaginary "$x_{\omega+1}$." What the solution to Thomson's Lamp above, together with the concomitant discovery and resolution of an apparent paradox between metrical and ordinal limits demonstrate, is precisely the stakes of the conceptual move at work in imagining $x_{\omega+1}$.

**CONCLUSION**

While Frege's identity of content sign explains *how* contact between the metrical and ordinal occurs in Equation 2, Deleuze's fold explains *what* that contact is. It is tempting to conclude a general philosophical schema for supertasks exists. Namely, that a supertasks folds together two limits that operate on sets whose cardinalities are separated by one degree.[29]

---

[28] "Between the two [rational] points A and B…there always remains the possibility for carrying out the right isosceles triangle, whose hypotenuse goes from A to B, and whose summit, C, determines a circle that crosses the straight line between A and B. The arc of the circle resembles a branch of inflection…that from an irrational number, at the meeting of the curved and straight lines, produces a point-fold" (*F*, 18).

The philosophical understanding of supertasks presented here may accelerate discovery of higher-cardinality supertask conditions, while the computational model of the STS provides future researchers a robust scaffolding to compose a hierarchical typology of supertask complexity.

While tantalizing, these directions remains for future research, particularly in collaboration with other specialists.[30] Indeed, the current article would not be possible without the research and cooperation of a multidisciplinary field of academics.[31] It is our hope that this article also provides a compelling case for the contributions philosophy, both analytic and continental, can make to multidisciplinary research, particularly in the burgeoning field of hypercomputation.

---

[29] In technical language, assuming the continuum hypothesis, sets whose aleph numbers are one apart. If one does not wish to assume the continuum hypothesis, this relationship can be recast as follows: sets for whom one set's cardinality is equivalent to the cardinality of the power set of the other. An interesting question presents itself: if a given supertask has some aleph number $x$ and we attempt to fold it into another set with aleph number $x+2$, can we say that the supertask actually completes $x+1$ many times?

[30] In addition to the earlier referenced "Turing Unbound," cf. Etesi, Gabor and Nemeti, Istvan. "Non-Turing computations via Malament-Hogarth space-times." International Journal of Theoretical Physics 41 (2002) 341-370.

[31] Additional heartfelt thanks for clear and consistent conversation to [REDACTED]

**Works Cited**

Bar-Elli, Gilead. "Identity in Frege's Begriffsschrift: Where Both Thau-Caplan and Heck are Wrong." Canadian Journal of Philosophy 36.3 (2006): 355-370.

Benacerraf, Paul. "Tasks, Super-Tasks, and the Modern Eleatics." Journal of Philosophy 59.24 (1962): 765-784.

Deleuze, Gilles transl. Conley, Tom. *The Fold: Leibniz and the Baroque. Minneapolis: University of Minnesota Press*, 1993.

Frege, Gottlob. *Begriffsschrift, a formula language, modeled upon that of arithmetic, for pure thought*. (1879). *From Frege to Gödel: a source book in mathematical logic, 1879-1931*. ed. Heijenoort, Jean. Cambridge: Harvard University Press, 1967.

Thomson, J.F.. "Tasks and Super-Tasks." Analysis 15.1 (1954): 1-13.

Thomson, J.F.. "Comments on Professor Benacerraf's Paper." *Zeno's Paradoxes*. Salmon, Wesley C., ed. Reprint. Hackett Publishing, 2001. 130-138.

**Works Referenced**

Bailey, D.H., Borwein, P.B., and Plouffe, S. "A New Formula for Picking off Pieces of Pi," Science News, v 148, p 279 (Oct 28, 1995).

Courant, Richard, and E.J. McShane. *Differential and Integral Calculus*. 2nd ed. Vol. 1. London and Glasgow: Blackie & Son Limited, 1937.

Etesi, Gabor and Nemeti, Istvan. "Non-Turing computations via Malament-Hogarth space-times." International Journal of Theoretical Physics 41 (2002) 341-370.

Jech, Thomas J., *Set Theory*. Third Millennium Edition. Springer, 2003.

Levinas, Emmanuel. *Totality and Infinity: An Essay on Exteriority*. The Hague: M. Nijhoff ;, 1979.

von Neumann, John. "On the introduction of transfinite numbers." (1923). *From Frege to Gödel: a source book in mathematical logic, 1879-1931*. ed. Heijenoort, Jean. Cambridge: Harvard University Press, 1967.

Welch, P.D.. "Turing Unbound: on the extent of computation in Malament-Hogarth spacetimes." British Journal for the Philosophy of Science 59.4 (Dec. 2008): 659-674.